\documentclass[prl,showpacs,amsfonts,amsmath,twocolumn,superscriptaddress,a4paper]{revtex4}
\usepackage{array,amsmath,amsfonts,amssymb,dsfont}
\usepackage{natbib}
\usepackage[T1]{fontenc}
\usepackage{ae}
\usepackage{color}
\usepackage{graphicx}

\begin{document}

\title{Preparation of Subradiant States using Local Qubit Control in Circuit QED}

\author{S.~Filipp}
\email{filipp@phys.ethz.ch} \affiliation{Department of Physics, ETH
Zurich, CH-8093
  Zurich, Switzerland}
\author{A. F. van Loo}
 \affiliation{Department of Physics, ETH Zurich, CH-8093
   Zurich, Switzerland}
 \author{M.~Baur}
 \affiliation{Department of Physics, ETH Zurich, CH-8093
   Zurich, Switzerland}
 \author{L.~Steffen}
\affiliation{Department of Physics, ETH Zurich, CH-8093 Zurich,
Switzerland}
\author{A.~Wallraff}
\affiliation{Department of Physics, ETH Zurich, CH-8093 Zurich,
Switzerland}


\renewcommand{\i}{{\mathrm i}}
\def\1{\mathchoice{\rm 1\mskip-4.2mu l}{\rm 1\mskip-4.2mu l}{\rm 1\mskip-4.6mu l}{\rm 1\mskip-5.2mu l}}
\newcommand{\ket}[1]{|#1\rangle}
\newcommand{\bra}[1]{\langle #1|}
\newcommand{\braket}[2]{\langle #1|#2\rangle}
\newcommand{\ketbra}[2]{|#1\rangle\langle#2|}
\newcommand{\opelem}[3]{\langle #1|#2|#3\rangle}
\newcommand{\projection}[1]{|#1\rangle\langle#1|}
\newcommand{\scalar}[1]{\langle #1|#1\rangle}
\newcommand{\op}[1]{\hat{#1}}
\newcommand{\vect}[1]{\boldsymbol{#1}}
\newcommand{\id}{\text{id}}
\newcommand{\red}[1]{\textcolor{red}{#1} }

\begin{abstract}
Transitions between quantum states by photon absorption or emission
are intimately related to symmetries of the system which lead to
 selection rules and the formation of dark states. In a circuit quantum
electrodynamics setup, in which two resonant superconducting qubits
are coupled through an on-chip cavity and driven via the common
cavity field, one single-excitation
state remains dark.  Here, we demonstrate that this dark state can be
excited using local phase control of individual qubit drives to change the
symmetry of the driving field.
 We observe that the dark
state decay via spontaneous emission into the cavity is suppressed, a
characteristic signature of subradiance. This local control technique could be used to prepare and study highly correlated quantum states of cavity-coupled qubits.
\end{abstract}

\maketitle

Symmetry properties of a quantum system interacting with
a radiation field provide information about possible transitions
within the system. 
Symmetry operations such as translation, rotation or reflection which
leave the system invariant, lead to
selection rules in molecular and solid-state systems
\cite{Ludwig1988}. For an ensemble of identical atoms, the
symmetry under permutation of particles allows only for transitions between symmetric
collective states \cite{Dicke1954, Gross1982}. 
Such highly-entangled Dicke-states, like the single-excitation W-state,  have attracted a lot of attention in the field of quantum information processing with trapped ions \cite{Haeffner2005}, optical photons \cite{Eibl2004} or superconducting qubits \cite{Neeley2010a} due to their robustness under decoherence \cite{Guehne2008} and particle loss \cite{Duerr2000,Briegel2001}. Moreover, collective states have been used for
quantum information storage in atomic memories 
\cite{Sangouard2011}. 
The particular symmetry of Dicke states affects also their
decay. Spontaneous emission can be enhanced for superradiant states
\cite{Dicke1954, Gross1982}, or inhibited, if the symmetry of the
state does not allow for the emission of a photon. This effect is
known as subradiance and is closely related to the concept of
decoherence-free subspaces, regions in Hilbert space which are not
affected by decoherence and therefore appealing for quantum information
processing \cite{Lidar2003}. Though theoretically well
studied
\cite{Crubellier1985,Crubellier1986},
subradiant states are difficult to realize and experimental evidence
of subradiance is rare \cite{Pavolini1985,DeVoe1996}. 

\begin{figure}[!t]
  \centering
  \includegraphics[width=86mm]{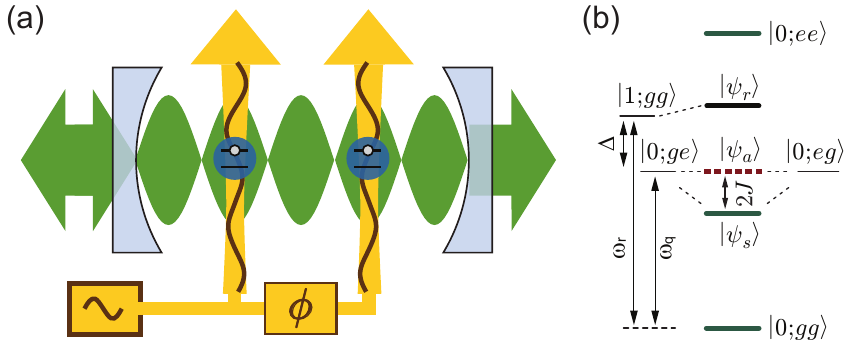}
  \caption{(a) Schematic drawing of a cavity QED setup with individual phase control of the driving field for each (artificial) atom. (b) Energy level diagram of the two-qubit system coupled dispersively to a common cavity field. Symmetric states are indicated by thick solid (green) lines and the antisymmetric state is identified by the thick dashed (red) line.}
  \label{fig:fig1}
\end{figure}

Here we present a method to prepare two qubits in antisymmetric
Dicke states and demonstrate their subradiance in circuit
 quantum electrodynamics (QED) \cite{Wallraff2004b,Blais2004}. In this
architecture, superconducting artificial atoms are coupled to a common
field mode of a planar microwave cavity. Strong resonant
coupling of individual qubits \cite{Houck2007, Fink2008, Bozyigit2010a} and qubit-ensembles \cite{Fink2009} to single microwave photons has been achieved. Moreover, cavity-mediated interactions between distant qubits \cite{Majer2007,Altomare2010,Filipp2011a} form the basis for on-chip quantum information processing
\cite{DiCarlo2009, Wang2010b} using entangled states of currently up to three qubits
\cite{DiCarlo2010}. In many experiments, the qubits are detuned
from the cavity resonance.  In this dispersive regime, the cavity can
 be used for the preparation and read-out of entangled states
\cite{Leek2009, Filipp2009b}, and cavity induced radiative decay of
the qubits due to the Purcell effect \cite{Purcell1946} is
reduced. This enhanced spontaneous emission can be modified by
 the qubit-cavity detuning \cite{Houck2008} and is the
dominant relaxation mechanism close to resonance. We have observed subradiance by preparing an antisymmetric
two-qubit state in this regime using local phase control to change the symmetry of the driving field [Fig.~\ref{fig:fig1}(a)]. 
 
We consider two qubits resonant with each
other but not with the cavity, modeled by a generalized Tavis-Cummings Hamiltonian \cite{Tavis1968}, 
\begin{align}
\label{eq:HTC}
H_{\rm{TC}} /\hbar &=  \omega_r a^\dagger a +  \omega_{q} J_z + g(a J_+ + a^\dagger J_-).
\end{align}
where the dominant cavity mode is at frequency $\omega_r$ and the
qubits are at frequency $\omega_q$.
The operators $J_z \equiv 1/2\sum_i^N \sigma_z^{(i)}$ and $J_\pm \equiv
\sum_i^N  \sigma^{(i)}_\pm$  are collective spin operators
\cite{Gross1982} with $\sigma^{(i)}_\pm \equiv (\sigma^{(i)}_x \pm i
\sigma^{(i)}_y)/2$ and Pauli
operators $\sigma^{(i)}_{x,y,z}$ for the individual qubits
($i=1,2$). $a^{(\dagger)}$ is the annihilation (creation)
operator of the field interacting with the qubits with equal coupling strength $g$. 
For the single-excitation manifold  the eigenstates of $H_{\rm{TC}}$ are
\begin{align}
\label{eq:states}
\ket{\psi_a} &= \ket{0;\psi_-},\\\nonumber
\ket{\psi_{r}} &= \cos\theta_m\ket{1;gg} + \sin\theta_m\ket{0;\psi_+},\\\nonumber
\ket{\psi_{s}} &= \sin\theta_m\ket{1;gg} - \cos\theta_m\ket{0;\psi_+},
\end{align} 
where the mixing angle $\theta_m$ is given by $\cos
2\theta_m = -\Delta/\sqrt{4(\sqrt{2}g)^2+\Delta^2}$. 
$\ket{n;}$ is a state with $n$ photons in the resonator, and 
$\ket{\psi_\pm} = (\ket{ge} \pm \ket{eg})/\sqrt{2}$ are the symmetric
and antisymmetric Bell-states. The qubit-resonator detuning $\Delta \equiv \omega_{q}-\omega_r $ is
chosen negative.  The antisymmetric state
$\ket{\psi_a}$ -- comprising the antisymmetric qubit state $\ket{\psi_-}$ -- does not couple to the cavity field. Only
the symmetric qubit state $\ket{\psi_+}$ mixes with the field to form hybridized matter-field states
$\ket{\psi_r}$ and $\ket{\psi_s}$ [Fig.~\ref{fig:fig1}(b)] with enhanced collective
coupling strength $\sqrt{N}g$ for $N=2$ \cite{Fink2009}. Generally, for an $N$-qubit ensemble there are always two hybridized
bright states  and $N-1$
uncoupled, dark states. Only qubit states which reflect the symmetry
of the cavity mode (i.~e. belong to compatible irreducible
representations of the permutation group \cite{Crubellier1985}) couple
to the single-photon field. In our case, the coupling to the first
harmonic cavity mode has the same sign for both
qubits ($g^{(1)}\approx g^{(2)}$) \cite{Filipp2011a}. Therefore,
the symmetric qubit state $\ket{\psi_+}$, which is invariant under
permutation of the qubits, couples to the field.

\begin{figure}[hb]
  \centering
  \includegraphics[width=86mm]{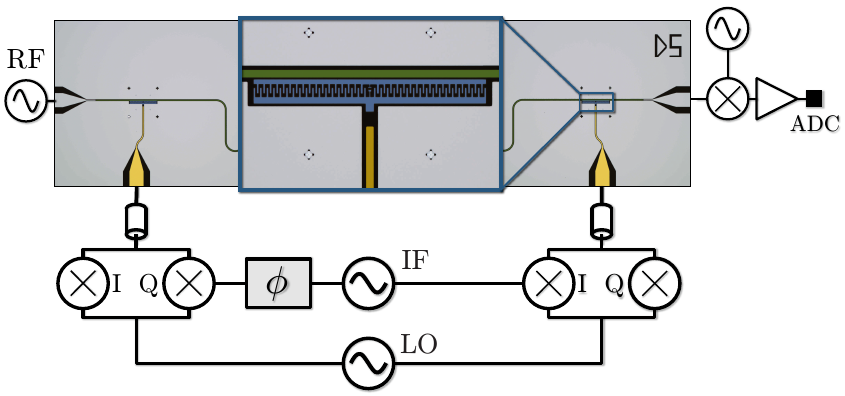}
  \caption{Setup and micrograph of the sample with a transmission line resonator (green) and two transmon qubits (blue) addressable via local charge lines (yellow).}
  \label{fig:fig2}
\end{figure}

For our experiments, two superconducting transmon qubits
\cite{Koch2007} have been integrated into a coplanar niobium resonator on a sapphire
substrate [Fig.~\ref{fig:fig2}(a,b)]. The qubits have similar Josephson energies $E_J/h \approx
 37.6~\rm{GHz}$, charging energies $E_C/h \approx 285~\rm{MHz}$ and coupling strengths $g/2\pi \approx 116~\rm{MHz}$ to the first
 harmonic mode of the microwave transmission line resonator. The resonator frequency is $\omega_r/2\pi = 6.937~\rm{GHz}$ and its decay rate is $\kappa/2\pi = 3.01~\rm{MHz}$.
In the dispersive regime where $\theta_m \approx \pi$, the photonic 
contribution $\ket{1;gg}$ to the symmetric
state $\ket{\psi_s}$ is small (of order $\sqrt{2}g/\Delta$) and the
state has pre-dominantly qubit character. It is however, shifted in
energy by $2 J \equiv 2 g^2/\Delta$ corresponding to the dispersive
$J$-coupling discussed in Refs.~\cite{Majer2007,Filipp2011a}
[Fig.~\ref{fig:fig1}(b)]. The antisymmetric wave-function
$\ket{\psi_a}$ has no photonic component and thus experiences no
Lamb-shift of its energy. In this description the qubit-qubit coupling
$J$ can be understood as the collective Lamb-shift $(\sqrt{2}
g)^2/\Delta = 2 J$ of the symmetric state $\ket{\psi_s}$.

The symmetry of these states is also reflected in selection rules for electric dipole transitions. In fact, for a drive applied directly to the cavity only, transitions from
the ground to the symmetric bright state $\ket{\psi_s}$ are allowed,
while transitions to the antisymmetric dark state $\ket{\psi_a}$ are
forbidden \cite{Filipp2011a}. Again, this is due to identical
microwave fields at the qubit positions for a
drive close to the first harmonic cavity mode, which conserves the
symmetry under permutation of qubits. Then only transitions within the
class of symmetric states [Fig.~\ref{fig:fig1}(b); solid green lines] are allowed \cite{Crubellier1985}.
This constraint can be overcome by addressing the qubits individually
via capacitively coupled charge lines \cite{Leek2009}
and tuning the relative phase $\phi$ of the microwave drive at
the qubit positions. When choosing a relative phase of $\phi=\pi$, the
opposite sign of the local fields results in allowed transitions to
the antisymmetric state \footnote{Recently, amplitude and phase control has also been used to
   suppress transitions in a two flux qubit system \cite{Groot2010b}.}.

In the rotating frame, the drive acting on the individual qubits with frequency $\omega_d$ and coupling 
$\epsilon$ is 
\begin{equation}
\label{eq:Hdrive}
H_d =  \hbar \epsilon \left(\sigma_+^{(1)} +
  \xi e^{i \phi}\sigma_+^{(2)}\right) + \emph{h.c.},
\end{equation}
 where $\xi$ is the amplitude imbalance at the individual qubits.
%
Starting in the ground state, the drive $H_d$ can induce transitions to the state $\ket{\psi}$, if the matrix element
$\Omega(\psi) \equiv |\langle 0;gg | H_d | \psi\rangle|/\hbar$
is non-zero. $\Omega(\psi_{\rm dark}) = 0$, therefore, defines
the dark state for given drive imbalance $\xi$ and relative
phase $\phi$. In the dispersive regime, the matrix element for the symmetric and antisymmetric state is 
\begin{equation}\label{eq:omega}\Omega(\psi_{s/a})=\epsilon\sqrt{(1 + \xi^2 \pm
  2\xi\cos\phi)/2}.
\end{equation}
 For equal drive amplitudes
($\xi=1$) and zero relative phase ($\phi = 0$), which corresponds to a drive applied to the cavity in the vicinity of the first harmonic mode, $\Omega(\psi_a)$ vanishes and the antisymmetric state $\ket{\psi_a}$ remains dark
\cite{Filipp2011a}. For
$\phi=\pi$ and $\xi =1$ however, $\Omega(\psi_a)$ is maximal, while  the transition rate
$\Omega(\psi_s)$ to the symmetric state vanishes. The transition can
thus be enabled, or disabled, by adjusting the relative phase appropriately.

\begin{figure}
  \centering
  \includegraphics[width=86mm]{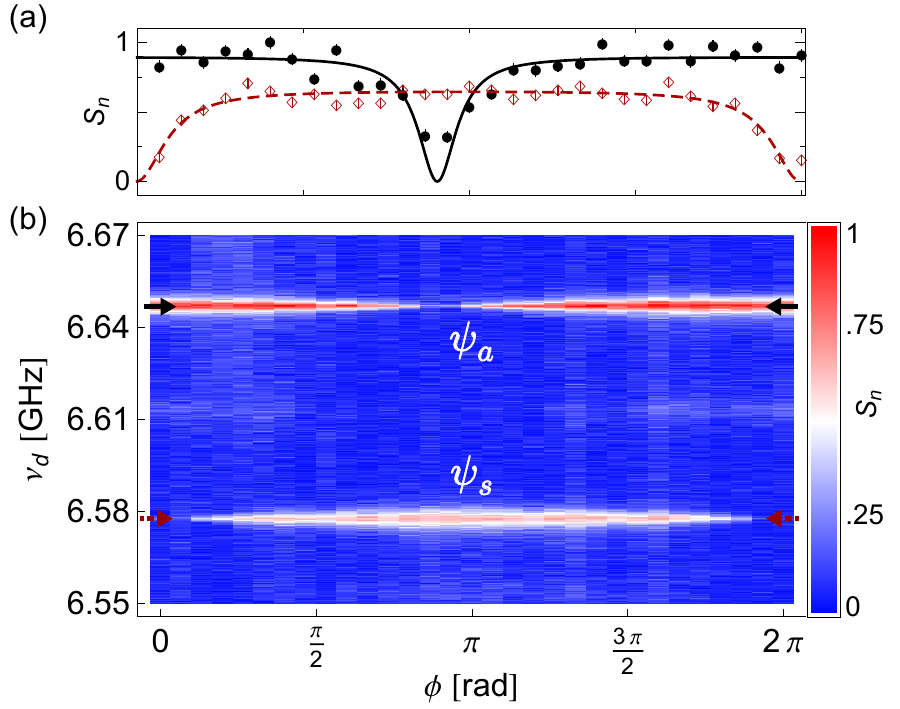}
  \caption{Spectroscopy of dark and bright state as a
    function of the relative phase $\phi$
    between drives. (a)
    Normalized integrated transmission amplitudes $S_{\rm{n}} = S/S_{\rm{max}}$ of the antisymmetric state $\psi_a$ (black solid dots) and the symmetric state $\psi_s$ (red open diamonds) at the frequencies $\omega_{a}/2\pi = 6.647~\rm{GHz}$ and $\omega_{s}/2\pi = 6.578~\rm{GHz}$, respectively, as indicated by the arrows in (b). $S_{\rm{n}}$ is normalized to unity at the maximum value of both curves $S_{\rm{max}}$. The lines indicate fits to the expected steady-state populations. (b) Transmission amplitude as a function of frequency $\nu_d$ and relative phase $\phi$.}
  \label{fig:fig3}
\end{figure}

We have spectroscopically measured the transition amplitude as a
function of relative phase $\phi$ in the vicinity of the bare qubit frequency
$\omega_{q}/2\pi = 6.647~\rm{GHz}$. To control $\phi$, the local
microwave fields are generated by a single microwave source operating
at the carrier frequency $\omega_{\rm{LO}} =
\omega_d+\omega_{\rm{IF}}$. We use two in-phase/quadrature (IQ) mixers to generate sidebands of
the carrier signal at the frequency $\omega_d$. The signal (IF) at the intermediate frequency
$\omega_{\rm{IF}} = 150~\rm{MHz}$ is synthesized with an arbitrary
waveform generator and applied  to each mixer with relative phase
$\phi$ [Fig.~\ref{fig:fig2}(a)]. The phase $\phi$ of the local qubit drive fields can then be controlled with high precision by the phase of the IF-signals. 
 Pulsed spectroscopy is used \cite{Wallraff2005,Bianchetti2009}, where the
measurement tone is switched on after a $500~\rm{ns}$ saturation pulse
[Fig.~\ref{fig:fig3}(a)]. The difference between the transmitted signal $s(t)$ and the signal with both qubits in their ground state $s_{gg}(t)$ integrated over
$t_m = 520~\rm{ns}$ yields the transmission amplitude $S = \int_0^{t_m} \{s(t)-s_{gg}(t)\} dt$. $S$ corresponds to a measurement of the steady-state population of
the single-excited qubit states when appropriately normalized
\cite{Filipp2009b}. 

Two spectroscopic lines, varying in amplitude as a function of relative phase $\phi$, are observed at $\omega_{a}/2\pi = 6.647~\rm{GHz}$ and $\omega_{s}/2\pi = 6.578~\rm{GHz}$. These correspond to the states $\ket{\psi_a}$ and $\ket{\psi_s}$, respectively. 
As expected, 
the zeros of the populations of the
symmetric and the antisymmetric state are 180$^\circ$ out-of-phase [Fig.~\ref{fig:fig3}].
The measured transmission amplitudes are in good agreement with the
steady-state population of the excited states calculated using the dissipative Bloch equations for a 
two-level system driven at zero detuning \cite{Abragam1961}, $S =
S_{0} \{1- 1/(1+ T_1 T_2 \Omega(\psi)^2)\}/2$.
The scaling factor $S_0$ takes account of the integrated voltage at the analog-digital converter. 
The dephasing times $T_{2,s} = 322\pm26~\rm{ns}$ and $T_{2,a} = 526\pm36~\rm{ns}$ of  $\ket{\psi_a}$ and $\ket{\psi_s}$  are determined in independent Ramsey-fringe experiments. 
The energy relaxation times $T_{1,s}$ and $T_{1,a}$ are measured in separate time-delay experiments and are discussed below.
 From
independent fits of $S$ to both curves we obtain the relative phase
difference between the zeros of the populations of the
symmetric and the antisymmetric state
$\phi_s-\phi_a = \pi-0.3\pm0.03~\rm{rad}$ using 
$\Omega(\psi_{s/a})$ from Eq.~(\ref{eq:omega}).
The deviation of the measured phase difference from $\pi$ is simply caused by
the difference in cable lengths of the two qubit drive lines. 

Using this method, it is possible to verify the subradiant character of the antisymmetric state by testing its resilience to
cavity-induced Purcell decay \cite{Purcell1946}, which is 
caused by the indirect coupling of the qubits to the environment via
the cavity. According to Fermi's golden rule, the induced voltage
fluctuations $\propto (a^\dagger + a)$ of the cavity field lead to a
decay rate $\gamma_\kappa = \kappa|\bra{0;gg} a \ket{\psi}|^2$ to the
ground state \cite{Koch2007,Houck2007}.  The total decay rate is then
given by $\gamma = \gamma_i + \gamma_\kappa$ with the intrinsic, non-radiative decay rate $\gamma_i$. Although the Purcell decay can be made small by operating the qubits in the dispersive regime, where $\gamma_\kappa\approx (g/\Delta)^2\kappa$, or by using advanced circuit designs \cite{Reed2010}, it cannot be fully avoided for single transmon qubits. 
For the dark state however, the matrix element $|\bra{0;gg} a
\ket{\psi_a}|$ vanishes completely, since by symmetry $\ket{\psi_a}$
has no photon admixture. In other words, destructive interference of
the photons emitted from either qubit leads to a suppression of the
spontaneous emission process and the dark state is protected against
Purcell decay. Note that a two-island transmon design, in
essence the integrated version of the two-qubit/cavity circuit used in
the experiments discussed here, also provides Purcell-protection based on the
formation of an intrinsic dark state \cite{Gambetta2011, Srinivasan2011}.

\begin{figure}
  \centering
  \includegraphics[width=86mm]{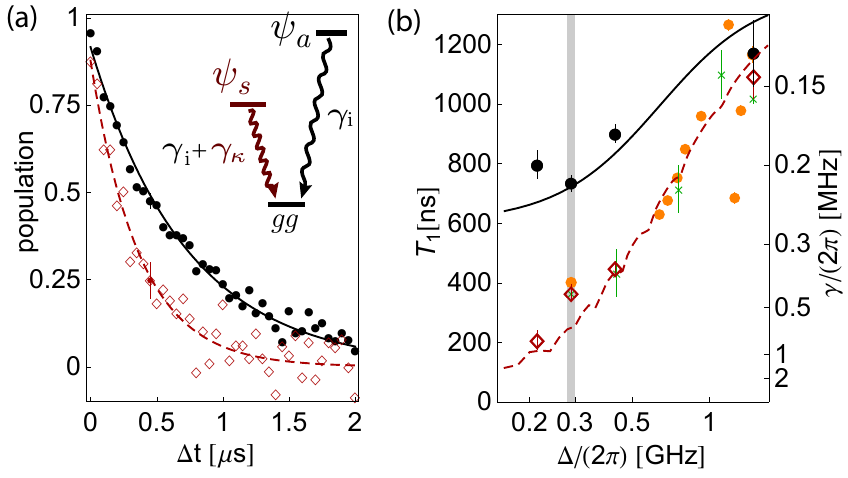}
  \caption{(a) Population of the antisymmetric dark ($\psi_a$, black dots) and the symmetric bright ($\psi_s$, red diamonds) state versus time at $\Delta/2\pi =
    290~\rm{MHz}$ (gray line in (b)) with exponential fits to the
    data. The inset shows the energy relaxation from $\psi_a$ and
    $\psi_s$ to the ground state, where only the symmetric state
    $\psi_s$ is affected by the Purcell decay $\gamma_\kappa$. (b)
    Measured decay times of $\psi_s$ (red diamonds), $\psi_a$ (black dots) and the uncoupled qubit
    states $eg$ (orange points) and $ge$ (green crosses) as a function
    of detuning $\Delta$. Exponential fits to numerically simulated
    populations are shown for $\psi_s$ (dashed line) and
    $\psi_a$ (solid line).}
  \label{fig:t1}
\end{figure}

In order to
observe subradiant Purcell-protection in a regime where radiative losses dominate over 
intrinsic qubit losses ($\gamma_\kappa > \gamma_i$), we have detuned
the qubits from the first harmonic mode by $\Delta/2\pi = 290~{\rm MHz}\sim 2.5 g$.
 At this frequency, the lifetime of the symmetric and antisymmetric state, as well as the decay
rates of the individual qubits  have been measured. A delayed measurement pulse
technique has been employed, where we apply a $\pi$ pulse resonant with the respective transition frequency and delay the time $\Delta t$ before applying the read-out pulse. 
The lifetimes of single qubit excitations $\rm{T}_{1,ge} = 401\pm
16~\rm{ns}$ and $\rm{T}_{1,eg} = 364\pm 16~\rm{ns}$ at this frequency
are comparable to the bright state lifetime of  $\rm{T}_{1,s} = 368\pm
30~\rm{ns}$. In our measurements, the effect of superradiant decay is masked by the
intrinsic decay rate and by dephasing acting locally on individual
qubits.
In contrast, the measured dark state lifetime $\rm{T}_{1,a} = 751\pm
13~\rm{ns}$ exceeds these values by a factor of two -- a clear
signature of subradiance that demonstrates the decoupling of the
antisymmetric state from the cavity environment and, as a consequence,
its enhanced stability. The population decay versus time of both the
bright and the dark state is plotted in Fig.~\ref{fig:t1}(a). 

The lifetime of the dark state is shown at different
detunings, along with the lifetimes of the bright state and the
uncoupled single qubit states in Fig.~\ref{fig:t1}(b). 
 It is verified in
numerical  master equation simulations  of the dissipative
dynamics for $\psi_a$ [Fig.~\ref{fig:t1}(b); solid black line] and
$\psi_s$ [Fig.~\ref{fig:t1}(b); dashed red line] that the decreasing lifetime 
of the dark state at small detunings is caused by local dephasing.
 For these simulation, a pure dephasing time of
$T_{2,\phi} = 880~\rm{ns}$ has been determined in Ramsey-fringe
experiments of the uncoupled qubits with an intrinsic decay time of
$1.37~\mu\rm{s}$  obtained from a fit to $\rm{T}_{1,eg}(\Delta)$.

The enhanced dark-state lifetime can be used for quantum
computation. In fact, the logical qubit formed by
 the ground and the dark state realizes a decoherence-free subspace,
 which is insensitive to cavity dissipation as well as to global dephasing acting
 simultaneously on both qubits. 
Note however, that the subspace spanned
by  $\ket{0;gg}$, $\ket{\psi_a}$ and the doubly excited state
$\ket{0;ee}$ forms a weakly nonlinear qubit with anharmonicity $2J$,
which limits the shortest preparation time without pulse optimization
\cite{Motzoi2009} to \mbox{$\sim\,1/(2J)$}.
 
More generally, this local control technique may allow us to excite highly entangled Dicke states belonging to different symmetry classes with a single microwave pulse conditioned on the choice of phases between individual drives. 
Moreover, the possibility to address states of different symmetry classes of multi-qubit systems
can be used to encode information in collective qubit
states. For readout, they can be transformed into entangled states in the computational basis by rapidly detuning the qubit transition frequencies.

In conclusion, we have demonstrated a method to  populate
dark states of a two-qubit system in a circuit QED setup. The transitions to either dark or bright two-qubit states
can be selected by adjusting the relative phase between individual
qubit drives, thus changing the
symmetry of the field and enforcing a symmetry-induced selection
rule. We apply this technique to demonstrate Purcell-protection of the
subradiant dark state against spontaneous emission. An extension to more qubits could provide further insight into the unitary and dissipative dynamics of multi-particle quantum states that can be directly prepared in the coupled qubit basis. Controlling the symmetry of the radiation field is, therefore, a viable method for preparation of states that are otherwise difficult to realize.

This work was supported by the Swiss National Science Foundation
(SNF). S.\,F. acknowledges support by the Austrian Science Foundation
(FWF). The authors thank A. Blais and J. Gambetta for valuable discussions.

\end{document}